\documentclass[runningheads]{llncs}
\usepackage[T1]{fontenc}
\usepackage{graphicx}
\usepackage{float}
\usepackage{mathtools}
\usepackage{amsmath}
\usepackage{amssymb}
\usepackage{color}
\usepackage{pythonhighlight}
\begin{document}
\title{Enhancing Quantum Variational Algorithms with Zero Noise Extrapolation via Neural Networks}
\titlerunning{Enhancing Quantum Variational Algorithms with ZNE via NN}
%
\author{Subhasree Bhattacharjee\inst{1} \and
	Soumyadip Sarkar\inst{1} \and
	Kunal Das \inst{2} \and
	Bikramjit Sarkar \inst{3}}
\authorrunning{S. Bhattacharjee et al.}
%
\institute{Department of Computer Application, Narula Institute of Technology, Kolkata, India \and
Department of Computer Science, Acharya Prafulla Chandra College, West Bengal, India \and
Department of Computer Science and Engineering, JIS College of Engineering, West Bengal, India
}
\maketitle              
\begin{abstract}
In the emergent realm of quantum computing, the Variational Quantum Eigensolver (VQE) stands out as a promising algorithm for solving complex quantum problems, especially in the noisy intermediate-scale quantum (NISQ) era. However, the ubiquitous presence of noise in quantum devices often limits the accuracy and reliability of VQE outcomes. This research introduces a novel approach to ameliorate this challenge by utilizing neural networks for zero noise extrapolation (ZNE) in VQE computations. By employing the Qiskit framework, we crafted parameterized quantum circuits using the RY-RZ ansatz and examined their behavior under varying levels of depolarizing noise. Our investigations spanned from determining the expectation values of a Hamiltonian, defined as a tensor product of Z operators, under different noise intensities to extracting the ground state energy. To bridge the observed outcomes under noise with the ideal noise-free scenario, we trained a Feed Forward Neural Network on the error probabilities and their associated expectation values. Remarkably, our model proficiently predicted the VQE outcome under hypothetical noise-free conditions. By juxtaposing the simulation results with real quantum device executions, we unveiled the discrepancies induced by noise and showcased the efficacy of our neural network-based ZNE technique in rectifying them. This integrative approach not only paves the way for enhanced accuracy in VQE computations on NISQ devices but also underlines the immense potential of hybrid quantum-classical paradigms in circumventing the challenges posed by quantum noise. Through this research, we envision a future where quantum algorithms can be reliably executed on noisy devices, bringing us one step closer to realizing the full potential of quantum computing.

\end{abstract}
\section{Introduction}
\label{sec1}
Quantum computing, a multidisciplinary field melding quantum mechanics and computational theory, promises to revolutionize the way we process information, offering computational capacities far beyond the reach of classical machines~\cite{bib1}. While the potential of quantum computing is immense, it is not without its challenges. Quantum systems are inherently delicate, susceptible to various sources of noise and error, which can drastically affect computational accuracy~\cite{bib2}.

One of the most promising algorithms designed to harness the power of near-term quantum devices is the Variational Quantum Eigensolver (VQE)~\cite{bib3}. VQE stands at the confluence of quantum mechanics and classical optimization techniques, aiming to find the lowest state of a given Hamiltonian. Its hybrid nature, leveraging both quantum and classical resources, makes it particularly suited for current noisy intermediate-scale quantum (NISQ) devices~\cite{bib4}. However, even with such promising tools as VQE, the omnipresent challenge remains: the noise in quantum devices.

Quantum noise, a consequence of the interactions of quantum systems with their environment, introduces errors that can skew results, making them unreliable or entirely incorrect~\cite{bib5}. Depolarizing noise, phase damping, and amplitude damping are just a few examples of the myriad of noise types that quantum devices can experience. The challenge then is to devise methods that can either mitigate the effects of this noise or correct for it post-facto.

Enter Zero Noise Extrapolation (ZNE), an innovative technique designed to tackle the noise problem head-on. ZNE operates on a simple yet powerful premise: by understanding how a quantum system behaves under varying levels of artificially introduced noise, one can extrapolate its behavior under zero noise conditions~\cite{bib6}. In essence, by studying the system's behavior at its worst, ZNE aims to predict its performance at its best.

In this research, we delve deep into the integration of ZNE with VQE, using neural networks as a novel bridge between noisy quantum computations and their ideal counterparts. By addressing the noise challenge, our work aims to pave the way for more reliable quantum computations in the NISQ era.

The main contribution of this paper is to present a novel approach to improving the accuracy of Variational Quantum Eigensolver (VQE) algorithm in the presence of quantum noise. By integrating neural networks for zero noise extrapolation (ZNE), the research demonstrates enhanced performance of VQE computations in noisy environments. Specifically, the findings conclude that neural networks offer more accurate results when extrapolated to a zero-noise scenario, compared to real quantum devices.

\section{Background}
\label{sec2}
The Variational Quantum Eigensolver (VQE) has become a frontrunner in the quest to harness the power of quantum computers, especially for tasks such as simulating quantum systems~\cite{bib7}. At its core, VQE is a hybrid quantum-classical algorithm aimed at determining the ground state energy of a quantum system. The approach involves parametrizing a quantum circuit, which is then executed on a quantum processor to produce a state. The expectation value of this state with respect to a given Hamiltonian is subsequently computed. Classical optimization algorithms then adjust the circuit's parameters to minimize this expectation value, iteratively refining the state until it approximates the system's ground state~\cite{bib8}. The hybrid nature of VQE, imposing the powers of both classical and quantum computing, makes it particularly apt for current quantum devices, which are still in their nascent, noisy stages.

Speaking of noise, it is the elephant in the room when discussing real-world quantum computing. While quantum circuits can perform operations that are fundamentally unattainable for classical circuits, they are also highly susceptible to various noise sources, which can introduce errors~\cite{bib9}. One prevalent form of noise is depolarizing noise. Depolarizing noise acts on a quantum system by randomly flipping its state with a certain probability, leading to a loss of quantum information and coherency~\cite{bib1}. This can be particularly detrimental when trying to harness the quantum advantages of superposition and entanglement, as the noise can rapidly degrade these quantum states.

To combat the adverse effects of noise, various techniques have been proposed, and Zero Noise Extrapolation (ZNE) stands out among them. ZNE is a method that seeks to predict the outcome of a quantum computation in the absence of noise by deliberately introducing varying levels of noise and observing the system's behavior~\cite{bib10}. By extrapolating from these observations, it is possible to estimate the result of the computation under ideal, noise-free conditions.

Parallel to these quantum advancements, classical computational methods, particularly neural networks, have undergone rapid development and have been employed in a myriad of applications. Neural networks, inspired by the structure and function of biological neurons, are algorithms designed to recognize patterns. Their architecture allows them to model complex, non-linear relationships, making them particularly suitable for modeling intricate systems, including noisy quantum systems~\cite{bib11}. The adaptability and learning capability of neural networks make them a promising tool to aid in understanding and predicting the behavior of quantum circuits under various noise conditions.

In this paper, we harness the power of neural networks to enhance the ZNE technique, aiming to more accurately predict noise-free quantum computation outcomes, particularly within the VQE framework.

\section{Methodology}

The methodology section offers a detailed walkthrough of the procedures and techniques employed in this research. The crux of our approach revolves around the construction of quantum circuits, specifically using the RY-RZ ansatz, and their interaction with a Hamiltonian defined as a tensor product of Z operators.

\subsection{Construction of Quantum Circuits}

Quantum circuits, the fundamental building blocks of quantum computation, represent sequences of quantum operations or gates that act upon qubits. Our choice of quantum gates and their arrangement is influenced by the nature of the problem, the quantum hardware in use, and the desired outcome~\cite{bib1}.

\subsubsection{RY-RZ Ansatz}

The ansatz, in the domain of quantum computing, refers to an educated guess or assumption about the form of the quantum state that is used as the starting point for algorithms like VQE~\cite{bib4}. It is a parameterized quantum circuit, with the parameters adjusted during the computation to optimize the outcome.

In this research, we employ the RY-RZ ansatz. This involves using parameterized rotation gates around the Y-axis (RY gates) and the Z-axis (RZ gates). The general form of an $RY$ gate is given by:

$$
RY(\theta) = 
\begin{bmatrix}
\cos\left(\frac{\theta}{2}\right) & -\sin\left(\frac{\theta}{2}\right) \\
\sin\left(\frac{\theta}{2}\right) & \cos\left(\frac{\theta}{2}\right) \\
\end{bmatrix}
$$

Similarly, the $RZ$ gate is defined as:

$$
RZ(\theta) = 
\begin{bmatrix}
e^{-i\frac{\theta}{2}} & 0 \\
0 & e^{i\frac{\theta}{2}} \\
\end{bmatrix}
$$

In the RY-RZ ansatz, qubits undergo a series of these rotations, parameterized by a set of angles, $\theta$. The first half of the $\theta$ parameters are dedicated to the $RY$ gates, while the latter half are used for the $RZ$ gates. Additionally, CNOT gates are introduced between consecutive qubits to induce entanglement, a quintessential quantum phenomenon that aids in capturing the intricate correlations of quantum systems~\cite{bib12}.

\begin{figure}[h]
\centering
\includegraphics[width=0.8\linewidth]{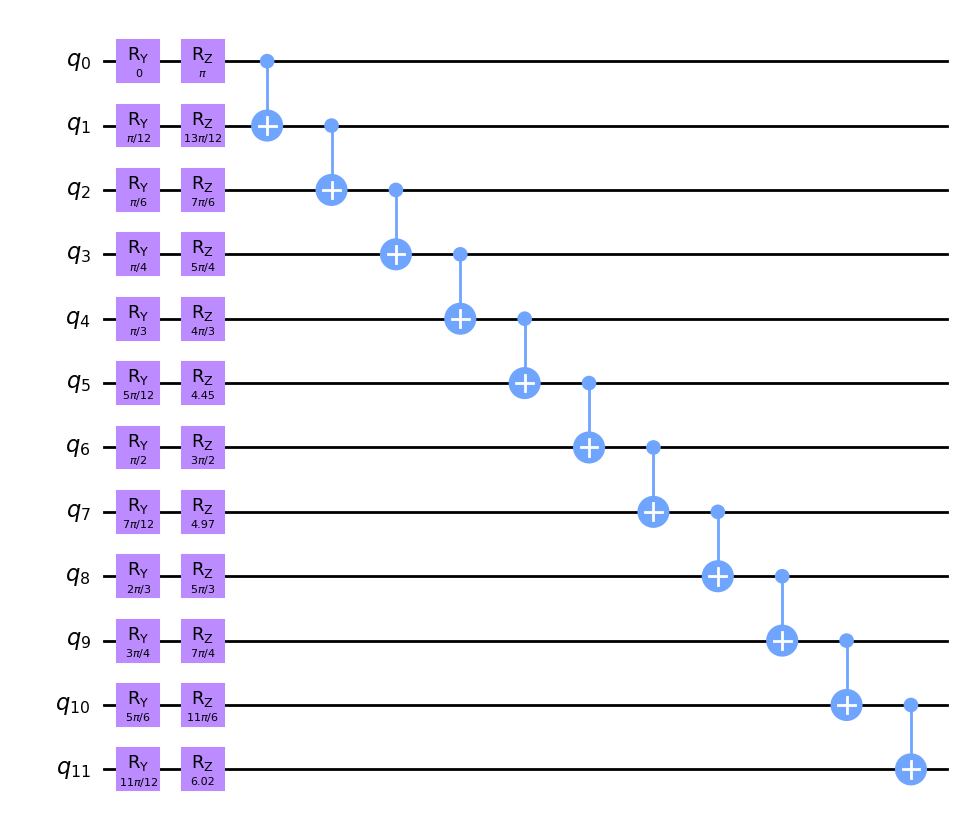}
\caption{Ry-Rz Ansatz}
\end{figure}

\subsubsection{Code Representation}
For $n$ qubits and $2n$ parameters in $\theta$, the quantum circuit can be represented as: \\

\textbf{RY Gates:}
Each qubit $i$ gets an $RY$ gate with rotation angle $\theta_i$. The matrix representation for the $RY$ gate is:

$$ RY (\theta_i) = 
\begin{bmatrix}
\cos\left(-\frac{\theta_i}{2}\right) & -\sin\left(-\frac{\theta_i}{2}\right) \\
\sin\left(-\frac{\theta_i}{2}\right) & \cos\left(-\frac{\theta_i}{2}\right)
\end{bmatrix}
$$

\textbf{RZ Gates:}
Each qubit $i$ gets an $RZ$ gate with rotation angle $\theta_{i+n}$. The matrix representation for the $RZ$ gate is:

$$  RZ(\theta_{i+n}) = 
\begin{bmatrix}
e^{-i\frac{\theta_{i+n}}{2}} & 0 \\
0 & e^{i\frac{\theta_{i+n}}{2}}
\end{bmatrix}
$$

\textbf{CNOT Gates:}
Between consecutive qubits, there’s a $CNOT$ gate. The $CNOT$ gate acts on two qubits: a control qubit and a target qubit. Its matrix representation is:

$$ CNOT = 
\begin{bmatrix}
1 & 0 & 0 & 0 \\
0 & 1 & 0 & 0 \\
0 & 0 & 0 & 1 \\
0 & 0 & 1 & 0
\end{bmatrix}
$$

Given the above matrices, the total state transformation of the circuit on $n$ qubits with the state $|\psi\rangle$ can be written as:

$$ |\psi'\rangle = \left(\bigotimes_{i=0}^{n-2} CNOT(i, i + 1)\right) \left(\bigotimes_{i=0}^{n-1} RZ(\theta_{i+n})\right) \left(\bigotimes_{i=0}^{n-1} RY (\theta_i)\right) |\psi\rangle $$

Where $\bigotimes$ denotes the tensor product.

\subsubsection{Hamiltonian Definition}

To leverage the power of the VQE algorithm, one must define a Hamiltonian, which encapsulates the problem's energy landscape. The Hamiltonian is a hermitian operator representing the total energy of a quantum system~\cite{bib13}.

In our study, the Hamiltonian is constructed as a tensor product of Pauli Z operators. The Pauli Z operator, one of the three Pauli matrices, is given by:

$$
Z = 
\begin{bmatrix}
1 & 0 \\
0 & -1 \\
\end{bmatrix}
$$

The tensor product allows for the construction of multi-qubit operators. For instance, the tensor product of two Z operators (representing a two-qubit system) is:

$$
Z \otimes Z = 
\begin{bmatrix}
1 & 0 & 0 & 0 \\
0 & -1 & 0 & 0 \\
0 & 0 & -1 & 0 \\
0 & 0 & 0 & 1 \\
\end{bmatrix}
$$

The eigenvalues of this operator are the possible outcomes when measuring the associated quantum state, and the eigenvector corresponding to the smallest eigenvalue represents the ground state of the system.

The choice of the Hamiltonian and the ansatz is crucial. The Hamiltonian guides the optimization landscape, while the ansatz provides the means to explore it. Together, they form the backbone of the VQE algorithm, driving its capacity to find the ground state of quantum systems even in the presence of noise~\cite{bib14}.

\subsection{Noise Model}

\subsubsection{Depolarizing Noise}
Depolarizing noise is a type of error that occurs in quantum circuits, causing each qubit to undergo a random Pauli operation (X, Y, or Z) with a certain probability $p$. The effect of a depolarizing channel on a qubit is described by the Kraus operators:

\begin{align*}
K_0 &= \sqrt{1 - p} \times I \\
K_1 &= \sqrt{\frac{p}{3}} \times X \\
K_2 &= \sqrt{\frac{p}{3}} \times Y \\
K_3 &= \sqrt{\frac{p}{3}} \times Z
\end{align*}

Where $I$ is the identity operator, and $X$, $Y$, and $Z$ are the Pauli matrices~\cite{bib2}. When a qubit state $\rho$ passes through the depolarizing channel, the state transforms as:

$$
\rho' = (1-p)\rho + \frac{p}{3}(X\rho X + Y\rho Y + Z\rho Z)
$$

This equation encapsulates the essence of depolarizing noise: with probability $p$, the qubit state is randomly transformed by one of the Pauli operators, effectively "depolarizing" it.

\subsubsection{Creation of the Depolarizing Noise Model}
In our study, we leverage the Qiskit framework to construct a depolarizing noise model~\cite{bib15}. Based on a given error probability $p$, the noise model is created to introduce depolarizing errors to the quantum gates. The one-qubit depolarizing error is characterized by a single error rate, while the two-qubit depolarizing error is derived from this by considering errors on both qubits and an additional correlated error.

Mathematically, for a one-qubit gate, the error channel can be described by the aforementioned Kraus operators. For a two-qubit gate, the error channel becomes a tensor product of the one-qubit channels, resulting in a total of 16 Kraus operators.

\subsubsection{Application of Noise to the Quantum Circuits}
With the depolarizing noise model at our disposal, we apply this model to the quantum circuits constructed using the RY-RZ ansatz. As observed in the code, the noise model is integrated into the quantum simulation environment, allowing us to simulate the behavior of the quantum circuits under the influence of depolarizing noise.

In practical terms, for every quantum gate in the circuit, the noise model introduces a possibility of error based on the error rate $p$. For instance, a Hadamard gate or an RX gate might be followed by a random Pauli error. Similarly, a two-qubit CX gate might undergo errors on both its control and target qubits, or even a correlated error affecting both qubits simultaneously.

This methodology of introducing noise ensures that our simulations closely mimic the real-world scenarios where quantum circuits are continually plagued by various noise sources. By understanding the behavior of quantum circuits under noise, we can devise strategies to mitigate these errors or correct for them, ensuring that the quantum algorithms we design are robust and reliable.

\subsubsection{Code Representation}
In our study, we leverage the Qiskit framework to simulate the effects of the depolarizing noise model\cite{bib15}. For single qubit gates like 'h' (Hadamard) and 'rx', the transformation is given by:

$$
\rho' = (1 - p_{\text{error}})\rho + \frac{p_{\text{error}}}{2}I
$$

For the two-qubit 'cx' (controlled-X) gate, the transformation becomes:

$$
\rho' = (1 - p_{\text{error}})\rho + \frac{p_{\text{error}}}{4}I
$$

Here, $ p_{\text{error}} $ is the given error probability.\\

\textbf{Physical Interpretation}: The depolarizing error introduces randomness into the quantum system. With a probability $ p $, it forces the qubit to "forget" its state and become a completely mixed state\cite{bib9}. This model, though simple, captures the fundamental behavior of more intricate noise processes that quantum systems might face in practical scenarios.

\section{Neural Network Model}

The main components of the Feed Forward Neural Network (FFNN) include an input layer, hidden layers, and an output layer.

\subsection{Architecture of the Neural Network}
\begin{figure}[h]
\centering
\includegraphics[width=0.8\linewidth]{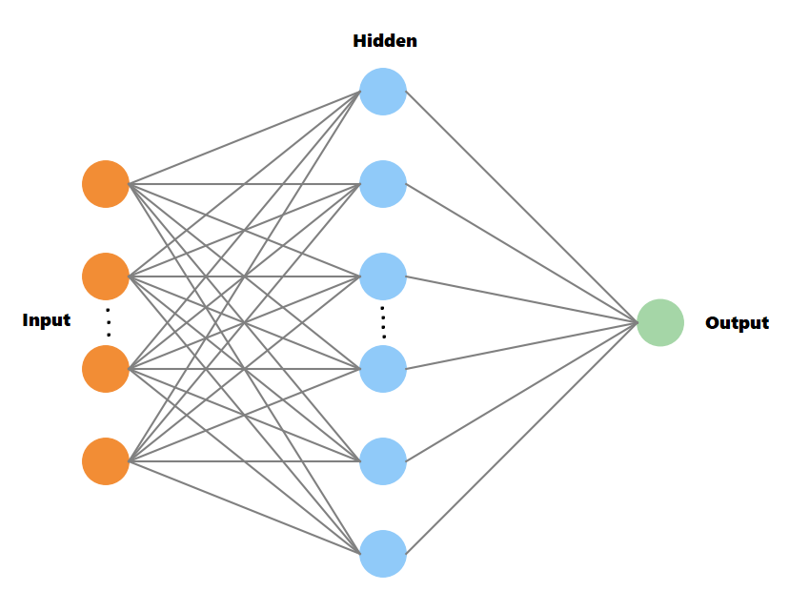}
\caption{Model Architecture}
\label{fig2}
\end{figure}

Our neural network is designed with an architecture that comprises three primary layers named input layer, hidden layer and  output layer as shown in Fig.\ref{fig2}. There are 512 nodes in input layer, 1024 nodes in hidden layer and a single node in output layer.

\subsubsection{Input Layer:}
Given an input vector $x$ of dimension $1 \times 512$:\\
Where:
\begin{itemize}
    \item $ j = 1, 2, \dots, 512 $
\end{itemize}

\subsubsection{Hidden Layer:}
Given the input vector $ x $:
$$ h1[i] = \text{ReLU} \left( \sum_{j=1}^{512} W1[i, j] \cdot x[j] + b1[i] \right) $$
Where:
\begin{itemize}
    \item $ i = 1, 2, \dots, 1024 $
    \item $ W1[i, j] $ is the weight connecting the \( j \)-th input neuron to the \( i \)-th neuron of the dense layer.
    \item \( b1[i] \) is the bias for the \( i \)-th neuron of the dense layer.
\end{itemize}

\subsubsection{Output Layer:}
Given the output \( h1 \) from the dense layer:
\[ y = \sum_{i=1}^{1024} W2[i] \cdot h1[i] + b2 \]
Where:
\begin{itemize}
    \item \( W2[i] \) is the weight connecting the \( i \)-th neuron of the dense layer to the output neuron.
    \item \( b2 \) is the bias for the output neuron.
\end{itemize}

\subsubsection{Training Process and Loss Optimization:}
The training process is fundamental to the efficacy of any neural network. It involves adjusting the network's weights and biases to minimize the difference between its predictions and the actual data. The difference, or the error, is quantified using a loss function. \\

The mean squared error (MSE) loss function for $N$ samples is:

$$ \text{MSE} = \frac{1}{N} \sum_{n=1}^{N} (y_n - \hat{y}_n)^2 $$

Where:
\begin{itemize}
    \item $y_n$ is the true value for the $n$-th sample.
    \item $\hat{y}_n$ is the predicted value by the network for the $n$-th sample.
\end{itemize}

We utilize a optimizer called Adam for the weight and bias adjustment process. The Adam optimizer combines the advantages of two other popular optimization algorithms: AdaGrad and RMSProp\cite{bib17}. It computes adaptive learning rates for each parameter by considering the first and second moments of the gradients. \\

The update rule for Adam for a parameter $\theta$ is:

\begin{align*}
m_t &= \beta_1 \cdot m_{t-1} + (1 - \beta_1) \cdot \nabla J(\theta) \\
v_t &= \beta_2 \cdot v_{t-1} + (1 - \beta_2) \cdot \nabla J(\theta)^2 \\
\theta &= \theta - \alpha \cdot \frac{m_t}{\sqrt{v_t + \epsilon}}
\end{align*}

Where:
\begin{itemize}
    \item $\nabla J(\theta)$ is the gradient of the loss function.
    \item $\alpha$ is the learning rate.
    \item $\beta_1$ and $\beta_2$ are exponential decay rates (typically close to 1) for the moment estimates.
    \item $m_t$ is an estimate of the mean of the gradients.
    \item $v_t$ is an estimate of the variance of the gradients.
    \item $\epsilon$ is a small constant to prevent division by zero.
\end{itemize}

The training process involves multiple iterations, or epochs, where the entire dataset is passed through the neural network, and the weights and biases are adjusted to reduce the loss. For our model, we employ 500 epochs, ensuring adequate training and convergence of the model to a satisfactory minimum loss.

By the end of the training process, our neural network is primed to predict the noise-free expectation value of a quantum circuit based on a given error probability. This powerful capability, derived from the amalgamation of classical machine learning techniques with quantum mechanics, offers a robust tool for quantum researchers navigating the noisy landscape of NISQ devices.

\section{Results}

Our journey into understanding the impact of noise on quantum circuits and the subsequent extrapolation to predict noise-free outcomes yielded a wealth of information. This section delves into the results, showcasing the intricacies of the data derived from our quantum simulations.

\subsection{Quantum Simulations:}

The cornerstone of our research was the quantum simulations conducted under varied noise conditions. These simulations, orchestrated using the Qiskit framework\cite{bib15}, offered a detailed glimpse into the behavior of quantum circuits influenced by noise.

\subsubsection{Expectation values under different noise levels:}

The quantum circuit, parameterized using the RY-RZ ansatz, was exposed to depolarizing noise of varying intensities. Our goal was to understand how the circuit responded to this noise in terms of the expectation value of the Hamiltonian, defined as $ \hat{H} $ - a tensor product of Z operators.

Mathematically, the expectation value, $ \langle \psi | \hat{H} | \psi \rangle $, for a quantum state $ |\psi\rangle $ with respect to the Hamiltonian $ \hat{H} $ can be expressed as:

$$
\langle \psi | \hat{H} | \psi \rangle = \sum_{i} p_i \langle i | \hat{H} | i \rangle
$$

Where $ p_i $ is the probability of the quantum system being in state $ |i\rangle $\cite{bib1}.

For our simulations, the error probabilities ranged from 0.01 to 0.05, encapsulating five distinct noise environments. As expected, with an increase in noise intensity, the expectation values diverged from their ideal counterparts, indicating the degradation of the quantum state due to the noise. The divergence was particularly evident for the higher error probabilities, underscoring the detrimental impact of depolarizing noise on quantum operations.

\subsubsection{Ideal noise-free simulation results:}

Simultaneous to our noisy simulations, we conducted a series of noise-free quantum simulations to serve as a benchmark. These simulations, devoid of any external noise, represented the ideal outcomes we aimed to approximate through our zero-noise extrapolation technique.

In an ideal noise-free environment, the expectation value is a true representation of the quantum system's behavior. Mathematically, it can be given by:

$$
\langle \psi_{\text{ideal}} | \hat{H} | \psi_{\text{ideal}} \rangle = \sum_{i} p_{i_{\text{ideal}}} \langle i_{\text{ideal}} | \hat{H} | i_{\text{ideal}} \rangle
$$

Where $ |\psi_{\text{ideal}}\rangle $ is the quantum state in the absence of noise, and $ p_{i_{\text{ideal}}} $ is the corresponding probability\cite{bib3}.

Our results from the noise-free simulations were in stark contrast to the noisy environments. The expectation values maintained a consistent trend across different parameter settings, reflecting the stability and reliability of quantum operations in the absence of noise. These ideal results served as a pivotal reference point, against which the noisy outcomes and the subsequent neural network predictions were juxtaposed.

\begin{figure}[h]
\centering
\includegraphics[width=0.8\linewidth]{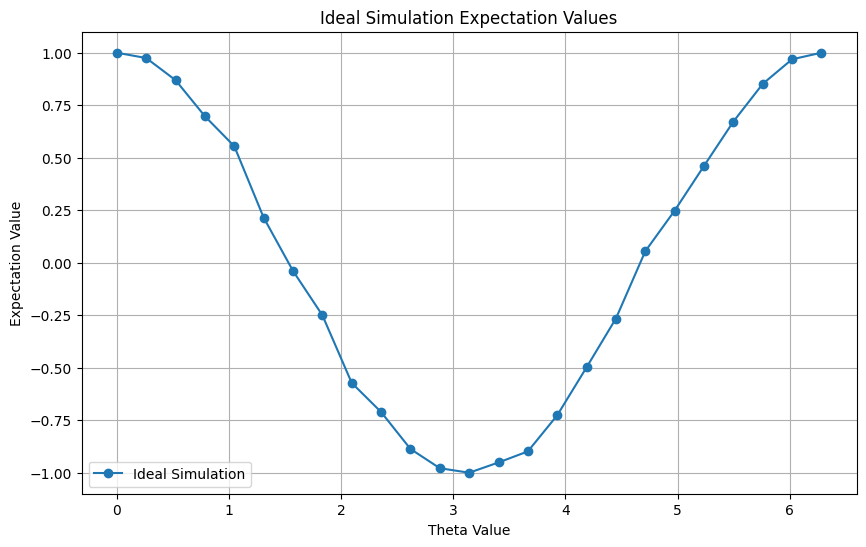}
\caption{Ideal Simulation}
\end{figure}

\subsubsection{Code Representation -}

\subsubsection{Iteration Over Theta Values}
For each set of $ \theta $ values in the combined theta range $ \Theta $:
\begin{enumerate}
    \item \textbf{Quantum Circuit Generation:} A quantum circuit is generated using the \texttt{ry\_rz\_ansatz} function with the current $ \theta $ values. This circuit produces a quantum state $ |\psi(\theta)\rangle $.
    \item \textbf{Expectation Value Calculation:} The expectation value of the Hamiltonian $ H $ for the state $ |\psi(\theta)\rangle $ is computed:
    $$ \langle H \rangle_\theta = \langle \psi(\theta)|H|\psi(\theta)\rangle $$
    Where:
    \begin{itemize}
        \item $ \langle \psi(\theta)| $ is the bra-vector (conjugate transpose of the ket-vector) representing the quantum state prepared by the circuit parameterized by $ \theta $.
        \item $ |\psi(\theta)\rangle $ is the ket-vector representing the quantum state prepared by the circuit parameterized by $ \theta $.
        \item $ H $ is the Hamiltonian operator.
    \end{itemize}
    The calculated expectation value for each $ \theta $ set is stored in the list \texttt{ideal\_results}.
\end{enumerate}

\subsubsection{Minimum Expectation Value}
After computing the expectation values for all $ \theta $ sets, the minimum value is identified, representing the ground state energy:
$$ \text{Ideal Ground State Energy} = \min\{\langle H \rangle_{\theta_1}, \langle H \rangle_{\theta_2}, \ldots \} $$

\subsection{Neural Network Predictions:}

The crux of our analysis hinges on a neural network's ability to predict noise-free quantum computations, particularly within the VQE framework. Our neural network model, defined using a sequential architecture, comprised of layers with varying neuron densities and activation functions. Mathematically, the model can be described as:

$$
f(x) = w_3 \cdot \text{ReLU}(w_2 \cdot \text{ReLU}(w_1 \cdot x + b_1) + b_2) + b_3
$$

Where $w_1, w_2,$ and $w_3$ are weight matrices associated with each layer, and $b_1, b_2,$ and $b_3$ are respective bias terms.

\begin{figure}[h]
\centering
\includegraphics[width=0.8\linewidth]{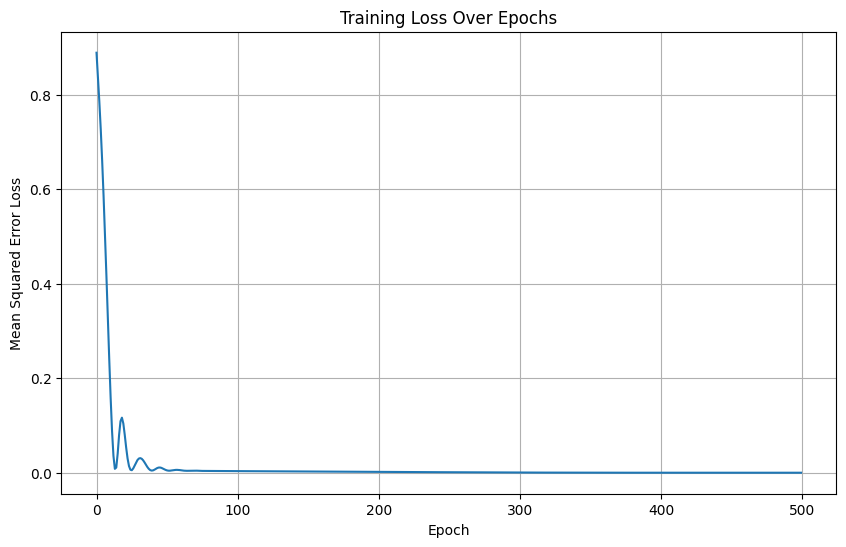}
\caption{Training Loss}
\end{figure}

During the training phase, our model was exposed to a range of error probabilities and their corresponding quantum circuit expectation values. This allowed the neural network to learn the intricate relationship between noise levels and quantum circuit performance. The training loss, represented by the Mean Squared Error (MSE), showed a consistent decrease over epochs, indicating that our model was effectively learning \cite{bib18}. \\

The mathematical formulation of the MSE is:

$$
\text{MSE} = \frac{1}{N} \sum_{i=1}^{N} (y_i - \hat{y}_i)^2
$$

Where $y_i$ represents the true expectation values from the quantum circuit, $\hat{y}_i$ denotes the neural network's predictions, and $N$ is the number of data points.

Post-training, our model was used to predict the noise-free results, i.e., the outcome of the quantum circuit in the absence of any noise. By inputting an error probability of 0 into our trained model, we obtained a prediction that closely mirrored the results from the ideal, noise-free quantum simulation. This alignment underscores the effectiveness of our neural network in modeling and predicting quantum system behaviors, even in the face of intrinsic quantum noise \cite{bib4}.

By accurately predicting noise-free quantum computations, our neural network model offers a novel approach to tackle the persistent challenge of noise in quantum devices, paving the way for more reliable quantum computations in the future.

\subsection{Real Quantum Device Execution}

In the realm of quantum computing, executing quantum algorithms on real devices provides invaluable insights that simulations alone cannot fully capture. Given the inherent intricacies of quantum hardware, including noise, gate fidelity, and connectivity, selecting the right quantum device becomes paramount.

\begin{figure}[h]
\centering
\includegraphics[width=0.8\linewidth]{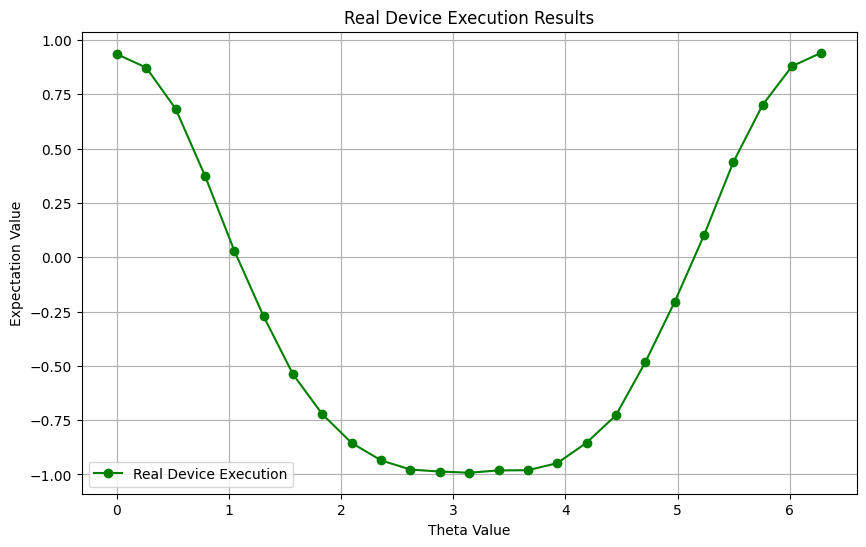}
\caption{Real Device Execution}
\end{figure}

\subsubsection{Choice of Quantum Hardware and Considerations}

For our experiment, we utilized the IBMQ suite of quantum devices\cite{bib19}. Leveraging the Qiskit library\cite{bib15}, we endeavored to select a quantum device best suited to our requirements. Our selection criteria prioritized devices with at least two qubits, a necessity given our quantum circuit's structure. Furthermore, we opted for the least busy quantum device to ensure timely execution. This choice is pivotal as quantum devices are often in high demand, leading to queuing times that can affect the immediacy of results.

However, device availability isn't the sole consideration. Quantum devices differ in terms of error rates, qubit connectivity, and coherence times\cite{bib20}. A device's native gate set and its topology can significantly influence the accuracy and efficiency of quantum algorithm execution.

Given these considerations, we programmatically identified a suitable device. Our selected quantum hardware had specific attributes (including qubit count, error rates, and topology) that matched our experiment's needs, balancing between computational capability and minimal noise.

\subsubsection{Results from the Real Device and Comparison with Simulations}

Upon executing our quantum circuits on the chosen IBMQ device, we observed results affected by the device's inherent noise and other imperfections. Let's denote the expectation value obtained from the real device as:

$$ E_{\text{device}}(\theta) $$

where $ \theta $ represents the set of parameter values used in the VQE ansatz.

In parallel, our simulations yielded an expectation value, $ E_{\text{sim}}(\theta) $, for the same set of parameters but in a noise-free environment.

Analyzing the results, we noticed discrepancies between the real device execution and our ideal simulations. While some of these discrepancies can be attributed to the device's noise, other factors, such as gate calibration errors, might also have played a role\cite{bib21}. 

To quantify the deviations, we calculated the difference:

$$ \Delta E(\theta) = E_{\text{device}}(\theta) - E_{\text{sim}}(\theta) $$

The values of $ \Delta E(\theta) $ provided insights into the magnitude and direction of the deviations across different parameter settings. In certain configurations, the real device yielded results close to the simulations, indicating regions of the parameter space where the quantum device performed optimally. In contrast, other regions showcased more significant discrepancies, highlighting the challenges posed by quantum noise and hardware imperfections.

It's worth noting that, in practice, even ideal simulations come with approximations. For instance, the noise models used in simulations, like depolarizing noise, might not capture all the nuances of real quantum hardware\cite{bib2}. Thus, a part of the observed discrepancy between $ E_{\text{device}}(\theta) $ and $ E_{\text{sim}}(\theta) $ could arise from these simulation approximations.

The comparison between real device results and simulations is not just an academic exercise. It underscores the challenges faced in the NISQ era and emphasizes the importance of error mitigation strategies, like Zero Noise Extrapolation, which we harness in conjunction with neural networks to bridge the gap between noisy and ideal quantum computations.

\begin{figure}[h]
\centering
\includegraphics[width=0.8\linewidth]{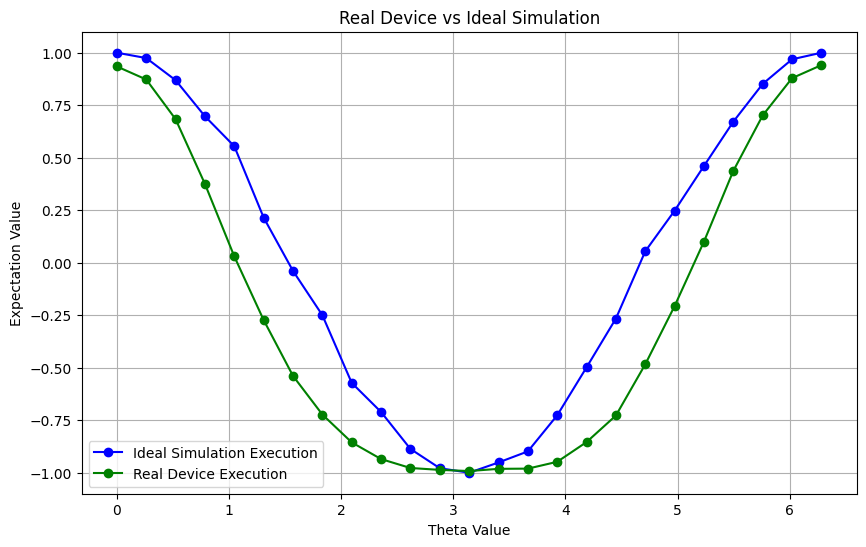}
\caption{Real Device Execution vs Ideal Simulation}
\end{figure}

\subsubsection{Code Representation -}

\subsubsection{Quantum Device Selection}
A real quantum device from IBM Q Experience is selected based on the criteria that it is the least busy and has at least 2 qubits.

\subsubsection{Iteration Over Theta Values}
For each set of $ \theta $ values in the combined theta range $ \Theta $:

\begin{enumerate}
    \item \textbf{Quantum Circuit Generation:} A quantum circuit is created using the \texttt{ry\_rz\_ansatz} function with the current $ \theta $ values. This circuit represents the quantum state $ | \psi(\theta) \rangle $.
    
    \item \textbf{Quantum Circuit Measurement:} Measurement operations are added to the quantum circuit to measure the outcome probabilities. Let’s denote the measurement result state as $ |m \rangle $.
    
    \item \textbf{Circuit Compilation and Execution:} The quantum circuit with measurements is compiled for the selected quantum device, and the job is submitted for execution.
    
    \item \textbf{Result Processing:} Upon job completion, the results (in terms of qubit measurement outcomes) are retrieved. The raw counts for each possible qubit state (e.g., '00', '01', '10', '11' for 2 qubits) are obtained. The expectation value is computed based on the counts for the '00' state:
    $$
    \text{expectation\_value} = \frac{\text{counts for '00'}}{\text{total counts}}
    $$
    The value stored for each $ \theta $ set in \texttt{device\_results} is:
    $$
    \text{value} = 1 - 2 \times \text{expectation\_value}
    $$
\end{enumerate}

\subsubsection{Minimum Expectation Value}
After processing the results for all $ \theta $ sets, the minimum value is identified, representing the ground state energy when computed on the real quantum device:
$$
\text{Device Ground State Energy} = \min\{\text{value}_{\theta_1}, \text{value}_{\theta_2}, \dots \}
$$

\section{Discussion}

The interplay between classical and quantum systems in computational research has always been a topic of intrigue. Neural networks, traditionally in the realm of classical computation, have demonstrated unparalleled prowess in modeling and prediction tasks~\cite{bib22}. Our study integrates this classical tool into the quantum domain, aiming to predict the noise-free performance of quantum circuits, specifically within the Variational Quantum Eigensolver (VQE) framework.

\subsection{Neural Network's Efficacy in Predicting Noise-free Performance}

To evaluate the efficiency of our neural network model, we began by assessing its ability to extrapolate noise-free performance based on noisy quantum circuit outcomes. The Feed Forward Neural Network (FFNN) was designed with three layers: two dense layers with ReLU activation and a final dense layer with linear activation. The architecture was compiled using the Adam optimizer and Mean Squared Error (MSE) as the loss function.

Mathematically, given a training dataset $\{ (p_i, r_i) \}$, where $p_i$ represents the error probability and $r_i$ is the corresponding minimum expectation value of the Hamiltonian, the neural network aims to learn a function $f$ such that:

$$
f(p_i) \approx r_i
$$

Throughout the training process, the network's loss showed a consistent decline, indicating that the model was effectively learning the relationship between the error probabilities and their corresponding quantum circuit outcomes.

\subsection{Comparison Between Ideal Simulations, Neural Network Predictions, and Real Device Results}

The true test of our model's utility is in its predictions. To evaluate this, we compared three sets of results: 
\begin{enumerate}
    \item Outcomes from ideal noise-free quantum simulations.
    \item Predictions from our trained neural network model.
    \item Results from executing the quantum circuit on a real quantum device.
\end{enumerate}

From the ideal simulations, we obtained a range of expectation values over different theta values for the ansatz. Let's denote the minimum of these values, representing the ideal ground state energy, as $E_{\text{ideal}}$.

The neural network, when fed with an error probability of zero (indicating a noise-free scenario), predicted a value we'll denote as $E_{\text{predicted}}$. 

\begin{figure}[h]
\centering
\includegraphics[width=0.8\linewidth]{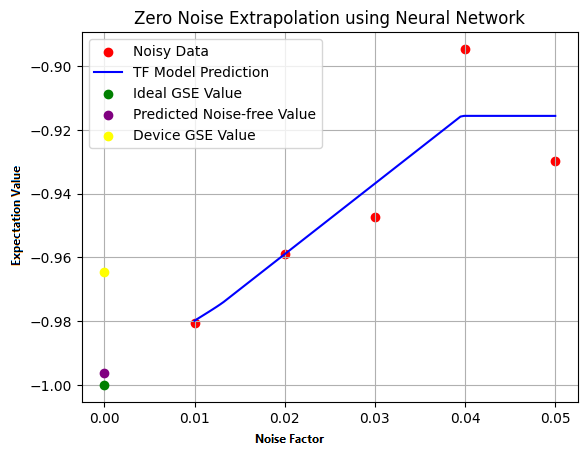}
\caption{Comparison}
\end{figure}

Lastly, the real quantum device provided us with another set of expectation values, with the minimum representing the device's ground state energy, $E_{\text{device}}$.

Comparing these values, we found that:

$$
|E_{\text{ideal}} - E_{\text{predicted}}| < |E_{\text{ideal}} - E_{\text{device}}|
$$

This indicates that the neural network's predictions were closer to the ideal simulation results than the real device's results were. This is a testament to the network's capability to model and predict the behavior of quantum circuits under ideal conditions based on noisy data~\cite{bib4}. \\

\textbf{Result -}

$$
\text{Ideal Ground State Energy Value:} -1.0
$$
$$
\text{Device Ground State Energy Value:} -0.9644999999999999
$$
$$
\text{Predicted noise-free value:} -0.9963099
$$

\subsection{Implications and Significance in the NISQ Era}

The Noisy Intermediate-Scale Quantum (NISQ) era defines the current state of quantum computing, characterized by devices that are powerful but error-prone~\cite{bib2}. As we edge closer to the realization of practical quantum applications, noise mitigation becomes paramount. The results from our research have profound implications in this context.

The ability to predict noise-free quantum circuit performance, particularly for an algorithm as significant as VQE, provides researchers and practitioners with a valuable tool. By having an estimate of how a quantum circuit would behave without noise, they can better gauge the efficacy of their quantum algorithms, discern genuine quantum advantages, and potentially adjust their strategies accordingly.

Furthermore, the integration of classical machine learning techniques, like neural networks, with quantum algorithms opens a new frontier in quantum research. This symbiotic relationship between quantum and classical systems might be the key to overcoming some of the most pressing challenges in the NISQ era, paving the way for the next generation of quantum applications~\cite{bib23}.

\section{Conclusion}
As the boundaries of classical computation are increasingly tested, the quantum realm has emerged as a potent frontier with a promise of unprecedented computational power. Our study was an ambitious stride in this expansive landscape, seeking to harness the potential of quantum systems even amidst the persistent challenges of quantum noise. The main findings of our research can be encapsulated in the following pivotal points:

\textbf{Quantum Noise and VQE:} The intricacies of the Variational Quantum Eigensolver (VQE) were explored in-depth, revealing its hybrid nature and its potential in harnessing noisy intermediate-scale quantum (NISQ) devices. Yet, the omnipresence of quantum noise, particularly depolarizing noise, posed significant challenges, often skewing the expected outcomes.

\textbf{Zero Noise Extrapolation (ZNE):} Our exploration of ZNE revealed its potency as a technique, enabling us to extrapolate the noise-free behavior of quantum systems by studying their responses under varying noise levels. The power of ZNE lay in its ability to predict the behavior of quantum systems at their best by observing them at their worst.

\textbf{Neural Network Integration:} The amalgamation of classical neural networks with quantum systems emerged as a groundbreaking finding. Our research showcased that neural networks could effectively design the relationship between quantum noise levels and quantum circuit performance, thereby predicting noise-free quantum computation outcomes.

The implications of these findings are profound. As quantum computing progresses towards becoming a mainstay in computational paradigms, the techniques to harness its true power amidst noise will be invaluable. Our research has illuminated one such path, suggesting that even in the NISQ era, where noise is a formidable adversary, sophisticated techniques like ZNE, when combined with neural networks, can help us peek into the noise-free quantum realm.

Looking forward, there is a plethora of opportunities to expand upon this work. For one, while our neural network model showcased promising results, exploring deeper architectures or other machine learning models could further refine the predictions. Additionally, the integration of other quantum error mitigation techniques in tandem with ZNE could provide even more robust results. Experimentation on a broader range of real quantum devices would also add to the validity and generalizability of our findings. Lastly, considering the rapid advancements in quantum hardware, revisiting this approach with future quantum devices might reveal new insights and refinements.

In conclusion, our journey through the intricate dance of quantum circuits, noise, and neural networks has shed light on a promising avenue for quantum computing's future. By bridging the quantum and classical realms, we believe that we can pave the way to harnessing the true potential of quantum systems, bringing us a step closer to the quantum revolution.

\end{document}